\documentclass[a4paper]{jpconf}
\usepackage{graphicx}

\newcommand{ \la }{\langle}
\newcommand{ \ra }{\rangle}
\newcommand{ \lp }{\left(}
\newcommand{ \rp }{\right)}
\newcommand{ \R }{{\cal R}}

\begin{document}
\title{Fluctuation and flow probes of early-time correlations in relativistic heavy ion collisions}

\author{Sean Gavin$^1$ and George Moschelli$^2$}

\address{$^1$~Department of Physics and Astronomy, Wayne State University, 666 W Hancock, Detroit,
MI, 48202, USA\\
$^2$~Frankfurt Institute for Advanced Studies, Johann Wolfgang Goethe University,
Ruth-Moufang-Str.~1, 60438 Frankfurt am Main, Germany}

\ead{$^1$~sean@physics.wayne.edu $^2$~moschelli@fias.uni-frankfurt.de}

\begin{abstract}
Fluctuation and correlation observables are often measured using multi-particle correlation methods and therefore mutually probe the origins of genuine correlations present in multi-particle distribution functions.  
We investigate the common influence of correlations arising from the spatially inhomogeneous initial state on multiplicity and momentum fluctuations as well as flow fluctuations. 
Although these observables reflect different aspects of the initial state, taken together, they can constrain a correlation scale set at the earliest moments of the collision. 
We calculate both the correlation scale in an initial stage Glasma flux tube picture and the modification to these correlations from later stage hydrodynamic flow and find quantitative agreement with experimental measurements over a range of collision systems and energies. 
\end{abstract}

%
%
%
%
\section{Introduction}
Fluctuating initial conditions of relativistic heavy ion collisions influence not only flow and but also fluctuation measurements. The random distribution of transverse parton density ``lumps" results in arbitrary event shapes.   Harmonic flow is generated largely by the global spatial anisotropy of the initial shape and induces a global correlation in the final momenta of measured hadrons. Conversely, the mere existence of parton density lumps results in local correlations; partons emerging from the same source are more likely to share the same phase space throughout the collision lifetime. We argue that the same local correlations that give rise to multiplicity and $p_t$ fluctuations drive fluctuations of the harmonic flow coefficients \cite{Gavin:2011gr,Gavin:2012if}.

To begin, in Sec. \ref{sec:cff} we describe multiplicity, $p_t$, and $v_n$ fluctuations as well as the ridge in terms of two-particle correlations. Using a cumulant expansion of the pair distribution reminiscent of those in Refs. \cite{Borghini:2000sa,Borghini:2001vi}, we show that all of the observables are linked by a common correlation function. The normalization of this correlation function sets a common scale that governs most of the centrality and center of mass collision energy dependence.

In Sec. \ref{sec:results} we investigate contributions to the correlation function from transversely local longitudinally long range correlations. 
We build on an approach started in Refs. \cite{Gavin:2008ev,Moschelli:2009tg} in which long range correlations result from the fragmentation of Glasma flux tubes. In this formulation local spatial correlations emerge from fluctuations in the number and distribution of flux tubes. These spatial correlations are then modified by transverse expansion, giving rise to azimuthal correlations. A similar physical picture motivates studies using a wide range of different techniques  \cite{Voloshin:2003ud,Pruneau:2007ua,Lindenbaum:2007ui,Sorensen:2008bf,Peitzmann:2009vj,Takahashi:2009na,Andrade:2010xy,Werner:2010aa}.

Finally in Sec. \ref{sec:discussion} we discuss how fluctuation measurements calculated in a unified framework can be used to constrain theories of the initial state of heavy ion collisions.
%
%
%
%
\section{\label{sec:cff}Correlations, Fluctuations, and Flow}
Correlation measurements largely rely on the distribution of hadron pairs $\rho_2(\mathbf{p}_1,\mathbf{p}_2) = dN/d\mathbf{p}_1d\mathbf{p}_2$.
Using the cumulant expansion method the pair distribution can be written as
\begin{equation}\label{eq:corrFunExp0}
	{\rho}_2(\mathbf{p}_1,\mathbf{p}_2) = 
	{\rho}_1(\mathbf{p}_1){\rho}_1(\mathbf{p}_2) +  r(\mathbf{p}_1, \mathbf{p}_2),    
\end{equation}
where $\rho_1(\mathbf{p}) = dN/d\mathbf{p}$ is the singles distribution, and $r(\mathbf{p}_1,\mathbf{p}_2)$ represents the genuine two-particle correlation function \cite{Gavin:2011gr,Gavin:2012if,Borghini:2000sa,Borghini:2001vi}. 

In this section we outline the role of correlations, $r(\mathbf{p}_1,\mathbf{p}_2)$, in determining fluctuation and flow observables and leave our model description to Sec. \ref{sec:results}. In the absence of correlations, the pair distribution factorizes into the square of the singles distribution, $ \rho_2(\mathbf{p}_1,\mathbf{p}_2) \rightarrow \rho_1(\mathbf{p}_1)\rho_1(\mathbf{p}_2)$. 
This assumption is often cited as a signature of harmonic flow, given that the event geometry is the dominant contributor to $v_n$ measurements. 
It is informative, however, to discuss the implications of relaxing this assumption when considering flow fluctuations. To address this we study other fluctuation observables such as multiplicity and momentum fluctuations which do not rely on event geometry but still reflect the existence of genuine correlations. We then find that although harmonic flow coefficients do depend on the global correlations induced by geometry, flow fluctuations emerge largely from the the same correlations as other fluctuation observables.
%
%
%
%
\subsection{\label{subsec:multfluc}Multiplicity Fluctuations}
\begin{figure}[h]
\includegraphics[width=14pc]{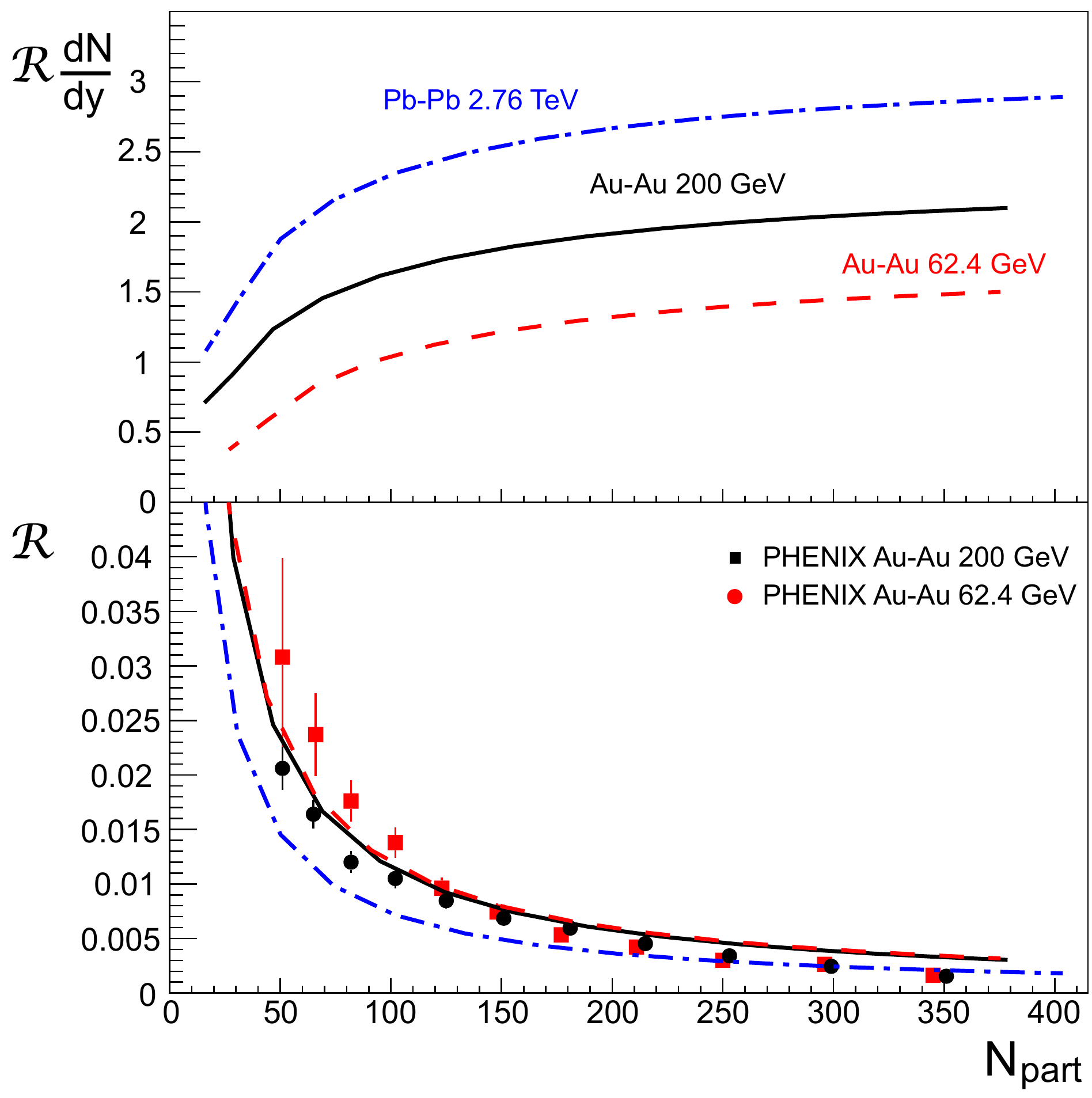}\hspace{2pc}%
\begin{minipage}[b]{14pc}\caption{\label{fig:Rscale}Prediction for ${\cal R}dN/dy$ as a function on the number of participants $N_{\rm part}$ at three beam energies (top). Calculated $\cal R$ compared to PHENIX data from \cite{Adare:2008ns} (bottom).}
\end{minipage}
\end{figure}

The existence of genuine correlations in the pair distribution signifies non-Poissonian particle production; all particles are not emitted from random independent sources \cite{Pruneau:2002yf}. Rearranging (\ref{eq:corrFunExp0}) as
\begin{equation}\label{eq:MomCorr0}
	r(\mathbf{p}_1, \mathbf{p}_2) = 
	{\rho}_2(\mathbf{p}_1,\mathbf{p}_2)-{\rho}_1(\mathbf{p}_1){\rho}_1(\mathbf{p}_2),    
\end{equation}
one can define the "robust variance",
\begin{equation}\label{eq:Rdef}
	{\cal R}  
	=\frac{1}{\langle N\rangle^2}
	\int r(\mathbf{p}_{1},\mathbf{p}_{2}) d\mathbf{p}_{1}d\mathbf{p}_{2}
	=\frac{\langle N(N-1)\rangle -\langle N\rangle^2}{\langle N\rangle^2}.
\end{equation}
We can rewrite the right most term of (\ref{eq:Rdef}) as $\R = Var(N)-\langle N\rangle/\langle N\rangle^2$. Possonian particle production requires $Var(N)=\langle N\rangle$, and consequently, non-zero values of $\R$ indicate a correlation in the production of hadrons. We will therefore refer to $\R$ as the correlation strength in this work. 
Significantly, the PHENIX collaboration measures multiplicity fluctuations using the width of the negative binomial distribution of produced hadrons \cite{Adare:2008ns}. The inverse of the width parameter is equivalent to the correlation strength, $\R=k^{-1}_{NBD}$ \cite{Gelis:2009wh}. The results are shown in the bottom panel of Fig. \ref{fig:Rscale}. This quantity sets the scale of all fluctuation measurements using two-particle correlations.

In Sec. \ref{sec:results} we will argue that these correlations originate from the fluctuating initial parton density. Pairs produced together in the hot spot or density lump are more likely to be correlated. We stress here that the shape of the event due to the distribution of hot spots does not contribute to multiplicity fluctuations.
%
%
Integration of (\ref{eq:Rdef}) over all momenta, including azimuthal angles, sums all correlated particle pairs, regardless of direction. 
%
%
%
%
\subsection{\label{subsec:ptfluc}Transverse Momentum Fluctuations}
Comparably to multiplicity fluctuations, many describe the dynamic fluctuations of transverse momentum using the covariance 
\begin{equation}\label{eq:ptFluctExp}
	\langle \delta p_{t1}\delta p_{t2}\rangle = 
	\frac{\langle \sum_{i \neq j}\delta p_{ti}\delta p_{tj}\rangle}{\langle N(N-1)\rangle},
\end{equation}
where $\delta p_{ti} = p_{ti}-\langle p_t\rangle$ and the average transverse momentum is $\langle p_t\rangle =  \langle P_t \rangle /\langle N\rangle$ for $P_t =  \sum_i p_{ti}$  the total momentum in an event \cite{Voloshin:1999yf,Voloshin:2001ei,Adamova:2003pz,Adams:2005ka}. This quantity vanishes when particles $i$ and $j$ are uncorrelated.  
%
%
Equation (\ref{eq:ptFluctExp}) sums pairs of differences in particle momenta from the average and can be written in terms of the pair correlation function (\ref{eq:MomCorr0}) as
\begin{equation}\label{eq:dptdpt}
	\langle \delta p_{t1}\delta p_{t2}\rangle =
	\int\! d\mathbf{p}_{1}d\mathbf{p}_{2}\,
	\frac{r(\mathbf{p}_{1},\mathbf{p}_{2})}{\langle N(N-1)\rangle}
	\delta p_{t1} \delta p_{t2}.
\end{equation}
Pairs  in which both particles have higher than average momentum add to  $\langle \delta p_{t1}\delta p_{t2}\rangle$. Lower-than-average pairs also add to the covariance, while high/low pairs subtract from it. In global equilibrium $\langle \delta p_{t1}\delta p_{t2}\rangle\equiv 0$. The presence of hot spots makes $\langle \delta p_{t1}\delta p_{t2}\rangle > 0$  (as would cold spots) \cite{Gavin:2003cb}. Motion of the sources would further enhance this quantity \cite{Voloshin:2003ud}. It follows that both jets and flow add to the $p_t$ covariance.

It is important to notice that momentum fluctuations depend on the same correlations as multiplicity fluctuations, the only difference being momentum weights. Like (\ref{eq:Rdef}) the integration in (\ref{eq:dptdpt}) counts pairs independent of their direction and therefore does not depend on the asymmetry of the system. Because of the momentum weights, (\ref{eq:dptdpt}) is sensitive to the average expansion and temperature and reflects the degree of thermalization of the system.
%
%
%
%
\subsection{\label{subsec:flowfluc}Flow Fluctuations}
Unlike multiplicity and momentum fluctuations, harmonic flow correlation and fluctuation measurements depend on the azimuthal anisotropy of the collision system. The anisotropy of the collision can be characterized by the moments 
\begin{equation}\label{eq:vnRP}
	\langle v_n\rangle = \langle \cos{n(\phi-\psi_{{}_{RP}})}\rangle,
\end{equation}
where $\langle ... \rangle$ represents an average over particles and events and $\psi_{_{RP}}$ is the reaction plane angle defined by the plane spanned by the impact parameter $\mathbf{b}$ and the beam direction. Since the reaction plane cannot be measured, experiments turn to the two-particle correlation measurement
%
%
%
%
\begin{equation}\label{eq:v2FluctExp}
	v_n\{2\}^2  = 
	\frac{\langle \sum_{i \neq j}\cos{n(\phi_i-\phi_j)}\rangle}
	{\langle N(N-1)\rangle}
\end{equation}
\cite{Borghini:2000sa,Borghini:2001vi,Bilandzic:2010jr,Voloshin:2008dg,Sorensen:2009cz}. Being a two-particle correlation, we can cast (\ref{eq:v2FluctExp}) in terms of the pair distribution 
\begin{equation}\label{eq:vn2dynamic}
	v_n\{2\}^2 =
	\int\! \,
	\frac{\rho_2(\mathbf{p}_{1},\mathbf{p}_{2})}{\langle N(N-1)\rangle}
	\cos{n(\phi_1-\phi_2)} \, d\mathbf{p}_{1}d\mathbf{p}_{2}.
\end{equation}
Using (\ref{eq:corrFunExp0}), (\ref{eq:vn2dynamic}) can be written as
\begin{equation}\label{eq:vnsigma}
	v_n\{2\}^2 = \langle v_n\rangle^2 + 2\sigma_n^2, 
\end{equation}
where the flow coefficient relative to the reaction plane is  $\langle v_n\rangle = \int \rho_1(\mathbf{p})\cos n(\phi-\Psi_{_{RP}})d\mathbf{p}$ comes from the single particle distribution and the factor of two in the fluctuation term $\sigma$ is conventional \cite{Voloshin:2007pc}. 
The effects of geometry primarily enter from the anisotropy of the single particle distribution characterized by (\ref{eq:vnRP}). 
Genuine correlations contributing to (\ref{eq:Rdef}) and (\ref{eq:dptdpt}) contribute to flow fluctuations
\begin{equation}\label{eq:sigmaDef}
	\sigma_n^2 =  
	\int \! d\mathbf{p}_{1}d\mathbf{p}_{2}\,
	\frac{r(\mathbf{p}_{1},\mathbf{p}_{2})}{2\langle N(N-1)\rangle}
	\cos n\Delta\phi.
\end{equation}
Ulike (\ref{eq:Rdef}) and (\ref{eq:dptdpt}), (\ref{eq:sigmaDef}) sums the relative azimuthal angles $\Delta\phi = \phi_1-\phi_2$ of correlated pairs. Anisotropic flow would modify correlated pairs emerging from common density sources since those pairs would experience a different push depending on their location in the transverse plane.

Thus far we have only mentioned correlations among flowing particles. The correlation function $r(\mathbf{p}_{1},\mathbf{p}_{2})$ also contains correlations emerging from jets, resonance decays, HBT effect, etc.. These correlations may or may not be modified by flow but contribute to all mentioned observables. To eliminate these effects we can consider long range correlations and measurements of observables with rapidity separations.
%
%
%
%
\subsection{\label{subsec:ridge}The Ridge}
Ridge measurements study two-particle azimuthal correlations at small and large (pseudo)rapidity separations $\Delta\eta$. Correlations at rapidity separations larger than $\sim 1-2$ units cannot come from jets, resonance decays or other short range phenomenon and causality dictates that such long range correlations originate in the early stages of the collision \cite{Dumitru:2008wn,Gavin:2008ev}. 

The soft ridge observable developed by STAR can be shown to relate directly to flow coefficients as well as flow fluctuations by 
\begin{equation}
	\frac{\Delta\rho(\Delta\phi,\Delta\eta)}{\sqrt{\rho_{\rm ref}}}
	= \frac{1}{2\pi}\frac{dN}{d\eta}\lp 2\sum\limits_{n=1}^{\infty} 
	\la v_n \ra^2 \cos n\Delta\phi
	+ \frac{r(\Delta\phi,\Delta\eta)}{\rho_{\rm ref}}\rp.
\end{equation}
\cite{Gavin:2012if}. 
Here the $\la v_n\ra$ terms follow from (\ref{eq:vnRP}), and $\rho_{\rm ref}$ is the experimental mixed-event background. We can isolate the contribution of flow fluctuations using
\begin{equation}\label{eq:FourierRidge}
	2\frac{dN}{dy}\sigma^2_{n} \approx
	\left. \int \frac{\Delta\rho(\Delta\phi,\Delta\eta)}
	{\sqrt{\rho_{\rm ref}}}\right|_{FS} 
	\cos (n\Delta\phi)~ d\Delta\phi d\Delta\eta,
\end{equation}
where $(\Delta\rho(\Delta\phi,\Delta\eta)/\sqrt{\rho_{\rm ref}})|_{FS}$ is the flow-subtracted ridge amplitude on the near side, centered at $\Delta\phi=\Delta\eta=0$  as reported by STAR \cite{Daugherity:2006hz,Daugherity:2008zz,Daugherity:2008su}.

Since ridge measurements rely on both flow and fluctuations, they can be compared to independent flow and flow fluctuation measurements to determine the relative contributions of long and short range correlations. Moreover, calculating all of the observables mentioned in Sec. \ref{sec:cff} within the same framework can constrain the common correlation function (\ref{eq:MomCorr0}) present in all measurements.
%
%
%
%
\section{\label{sec:results}Glasma Correlations and Results}
We argue that multiplicity, momentum, and flow fluctuations originate from initial state spatial correlations. These correlations emerge from fluctuating parton density distributions. Density lumps or hot spots emerge reflecting the positions of partons sources such as wounded nucleons or, in our case, Color Glass Condensate - Glasma flux tubes. We take correlated particles as those originating from the same source.

The spatial correlation function
\begin{equation}\label{eq:CorrDef}
	c(\mathbf{x}_1, \mathbf{x}_2) = 
	n_2(\mathbf{x}_1,\mathbf{x}_2) - n_1(\mathbf{x}_1)n_1(\mathbf{x}_2)
\end{equation}
identifies the initial distribution of correlated pairs in the transverse space. If the pair distribution, $n_2(\mathbf{x}_1,\mathbf{x}_2)$, contains no correlations, it factorizes into the square of the singles distribution, $n_1(\mathbf{x}_i)$ and (\ref{eq:CorrDef}) vanishes. 
We write (\ref{eq:CorrDef}) in a Glasma flux tube picture as
\begin{equation}\label{eq:CorrFunc}
	c(\mathbf{x}_1, \mathbf{x}_2)
	= {\la N\ra}^2\R\,\delta(\mathbf{r}_t) \rho_{_{FT}} (\mathbf{R}_t).
\end{equation}
The delta function in relative transverse elliptical coordinates $\mathbf{r}_t = \mathbf{r}_{t1} - \mathbf{r}_{t2}$ enforces the condition that correlated partons originate from the same flux tube presuming the flux tube transverse size is small compared to that of the collision area. 
The flux tube probability distribution, written in average coordinates $\mathbf{R}_t = (\mathbf{r}_{t1} + \mathbf{r}_{t2})/2$, is
\begin{equation}\label{eq:TubeDis}
	\rho_{_{FT}}(\mathbf{R}_t) \approx  
	\frac{2}{\pi R^2_A}  \lp 1 -\frac{R_t^2}{R_A^2}\rp.
\end{equation}
Here the shape of (\ref{eq:TubeDis}) resembles the nuclear thickness function and $\pi R_A^2$ is the area of the overlap region. The distribution (\ref{eq:TubeDis}) represents an average over all possible shapes, which we have taken to be a simple ellipse. 

To compute the correlation strength $\cal R$, we imagine each event produces $K$ flux tubes with transverse size $\sim Q_s^{-2}$. In the saturation regime $K$ is proportional to the transverse area $R_A^2$ divided by the area per flux tube, $Q_s^{-2}$ \cite{Kharzeev:2000ph}. Allowing $K$ to fluctuate from event to event with average $\la K \ra$, we calculate in Ref. \cite{Gavin:2008ev} that
\begin{equation}\label{eq:GlasR}
	{\cal R}
	=\frac{\la N^2 \ra -\la N \ra^2 -\la N \ra}{\la N \ra^2}
	\propto \langle K\rangle^{-1}.
\end{equation}
%
%
%
%
%
%
%
%
\begin{figure}[h]
\begin{minipage}{14pc}
\includegraphics[width=14.3pc]{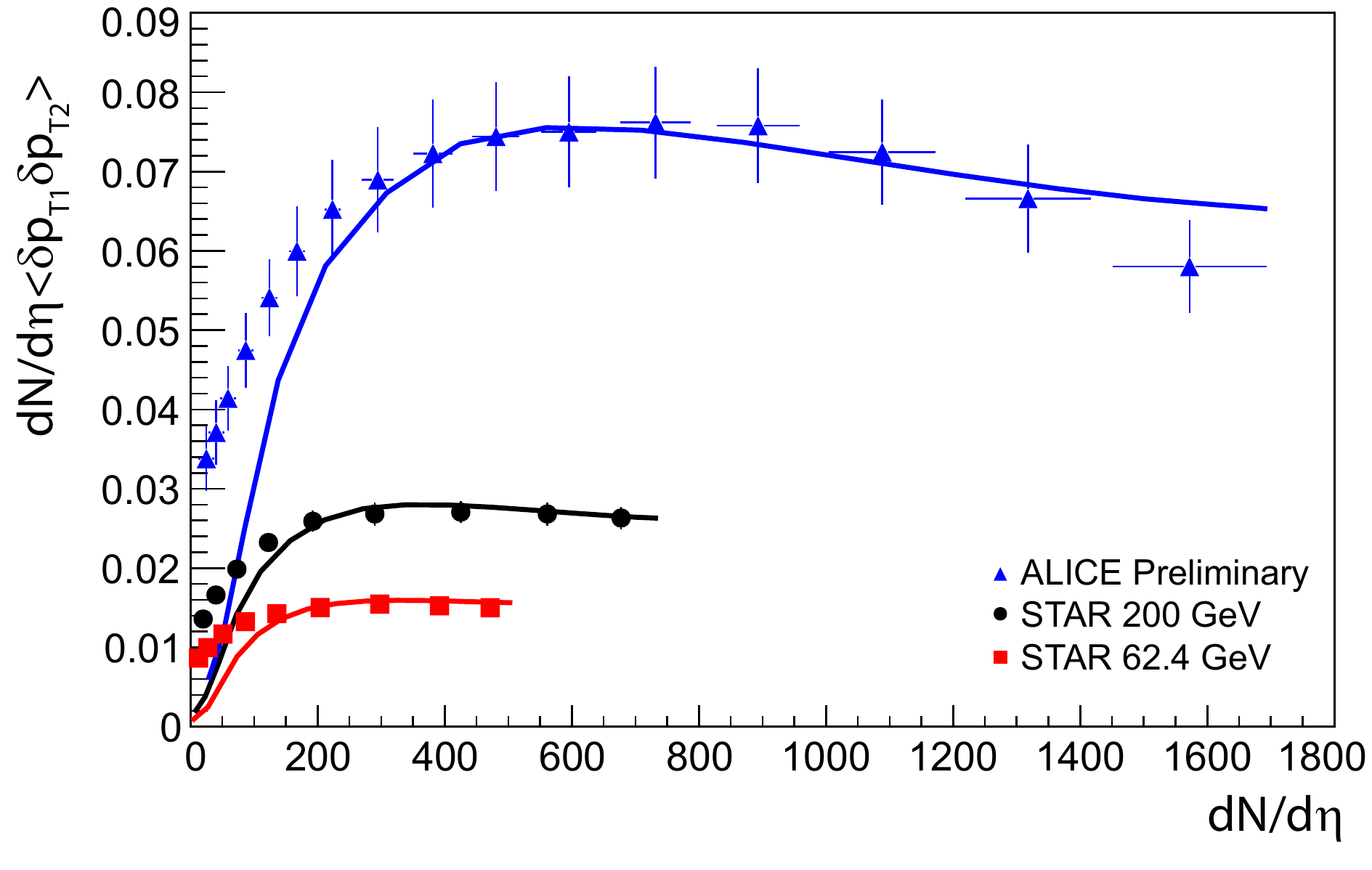}
\caption{\label{fig:dptdpt} Transverse momentum fluctuations $\langle \delta p_{t1}\delta p_{t2}\rangle dN/d\eta$ as a function on the number of participants $N_{\rm part}$ at three beam energies. Data is from \cite{Adams:2005ka,Heckel_ALICEptFluc}.}
\end{minipage}\hspace{2pc}%
\begin{minipage}{14pc}
\includegraphics[width=14pc]{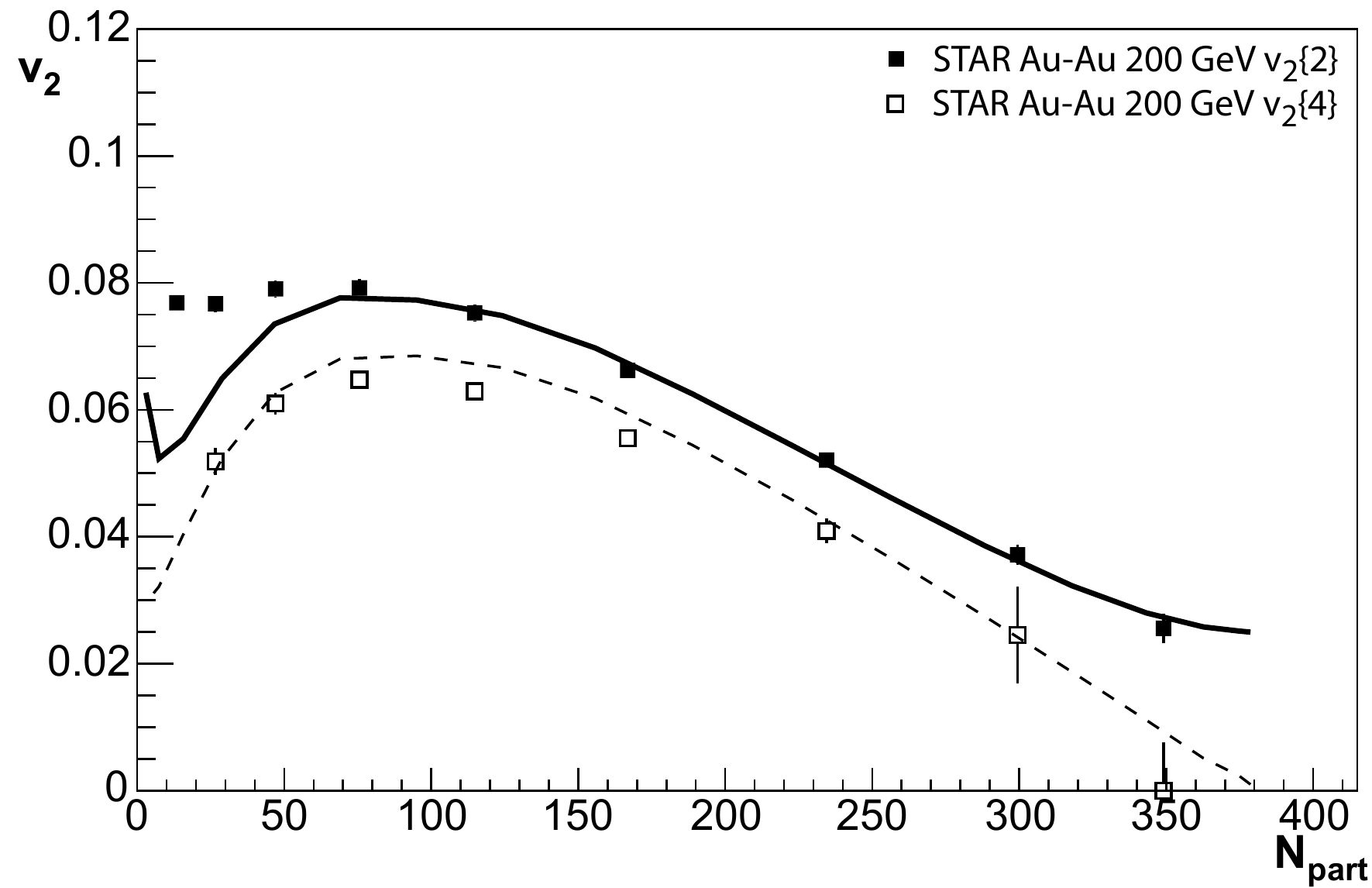}
\caption{\label{fig:v2STAR} Measured $v_2\{2\}$ and $v_2\{4\}$  from STAR \cite{Pruthi:2011eq} compared to calculations. 
The difference between the calculated curves is due to flow fluctuations.}
\end{minipage} 
\end{figure}
Each Glasma flux tube yields an average multiplicity of  $\sim \alpha_s^{-1}(Q_s)$ gluons and as in Ref. \cite{Kharzeev:2000ph}, the number of gluons in a rapidity interval $\Delta y$ is then
\begin{equation}\label{eq:Nscale}
	\langle N\rangle = 
	({{dN}/{dy}})\Delta y  \sim {\alpha_s}^{-1}(Q_s)\langle K\rangle.
\end{equation}
Finally, the Glasma correlation scale,
\begin{equation}\label{eq:CGCscale}
	\R{{dN}/{dy}} =
	\kappa {\alpha_s^{-1}}(Q_s^2),
\end{equation}
follows from the multiplication of (\ref{eq:Rdef}) and (\ref{eq:Nscale}). Notice
(\ref{eq:CGCscale}) is dimensionless and depends only on the saturation scale, $Q_s^2$, which can be calculated from first principles. Measurements of the ridge at various beam energies, target masses, and centralities fix the dimensionless coefficient $\kappa$ and are in excellent accord with the leading-order dependence  \cite{Gavin:2008ev,Moschelli:2009tg}. 

Intrinsically correlated partons originate from the same transverse position and experience, on average, the same transverse momentum modulation from flow. 
We represent the genuine two-particle momentum space correlation as
\begin{equation}\label{eq:MomCorr}
	r(\mathbf{p}_1, \mathbf{p}_2) =
	\!\!\int c(\mathbf{x}_1, \mathbf{x}_2) 
	f(\mathbf{x}_1,\mathbf{p}_1)
	f(\mathbf{x}_2,\mathbf{p}_2)
	d\Gamma_1d\Gamma_2
\end{equation}
We evaluate (\ref{eq:MomCorr}) using a blast-wave description of the Boltzmann phase-space density $f(\mathbf{x},\mathbf{p})$ on a Cooper-Frye freeze out surface $d\Gamma$ \cite{Gavin:2011gr,Gavin:2012if}. 
Notice that full integration of (\ref{eq:MomCorr}) yields $\la N\ra^2\R$ as does (\ref{eq:CorrDef}), tying azimuthal correlations to multiplicity and transverse momentum fluctuations \cite{Gavin:2011gr}.

Using (\ref{eq:MomCorr}) we calculate (\ref{eq:dptdpt}) and find the results shown in Fig. \ref{fig:dptdpt}. We find good agreement in central collisions where the assumption of gluon saturation necessary for the production of Glasma flux tubes is most valid. Moreover, the onset of thermalization in peripheral collisions should modify the correlation function.  In particular, this effect has been shown to modify $\langle \delta p_{t1}\delta p_{t2}\rangle$ at low numbers of participants \cite{Gavin:2003cb}. Partial thermalization describes peripheral RHIC data very well and, moreover, allows one to describe $pp$ and $AA$ collisions in the same model \cite{Gavin:2004dc}. 

Just as with (\ref{eq:dptdpt}) we can use the same correlation function (\ref{eq:MomCorr}) to study flow fluctuations (\ref{eq:sigmaDef}) and therefore examine the effects of event anisotropy. If flow fluctuations arise only from changes in the shape of the collision area, then their governing correlation function would predict $\langle \delta p_{t1}\delta p_{t2}\rangle = 0$ and have almost no dependence on collision energy. To start we calculate (\ref{eq:sigmaDef}) for 200 GeV Au+Au collisions. Using this along with the experimental definition
\begin{equation}\label{eq:sigvn}
\sigma_{n}^2 = \frac{v_n\{2\}^2 - v_n\{4\}^2}{2}.
\end{equation}
and taking $v_n\{4\}\approx\langle v_n\rangle$ we calculate $v_2\{2\}$ as the solid line in Fig. \ref{fig:v2STAR}. 
%
%
%
%
The dashed line is (\ref{eq:vnRP}) calculated from blast wave and flow fluctuations (\ref{eq:sigmaDef}) determine the difference between the two curves. Notice that in central collisions the circular event geometry drives $\langle v_2\rangle$ to zero. Here flow fluctuations are largest and completely determine $v_2\{2\}$.
\begin{figure}[h]
\includegraphics[width=14pc]{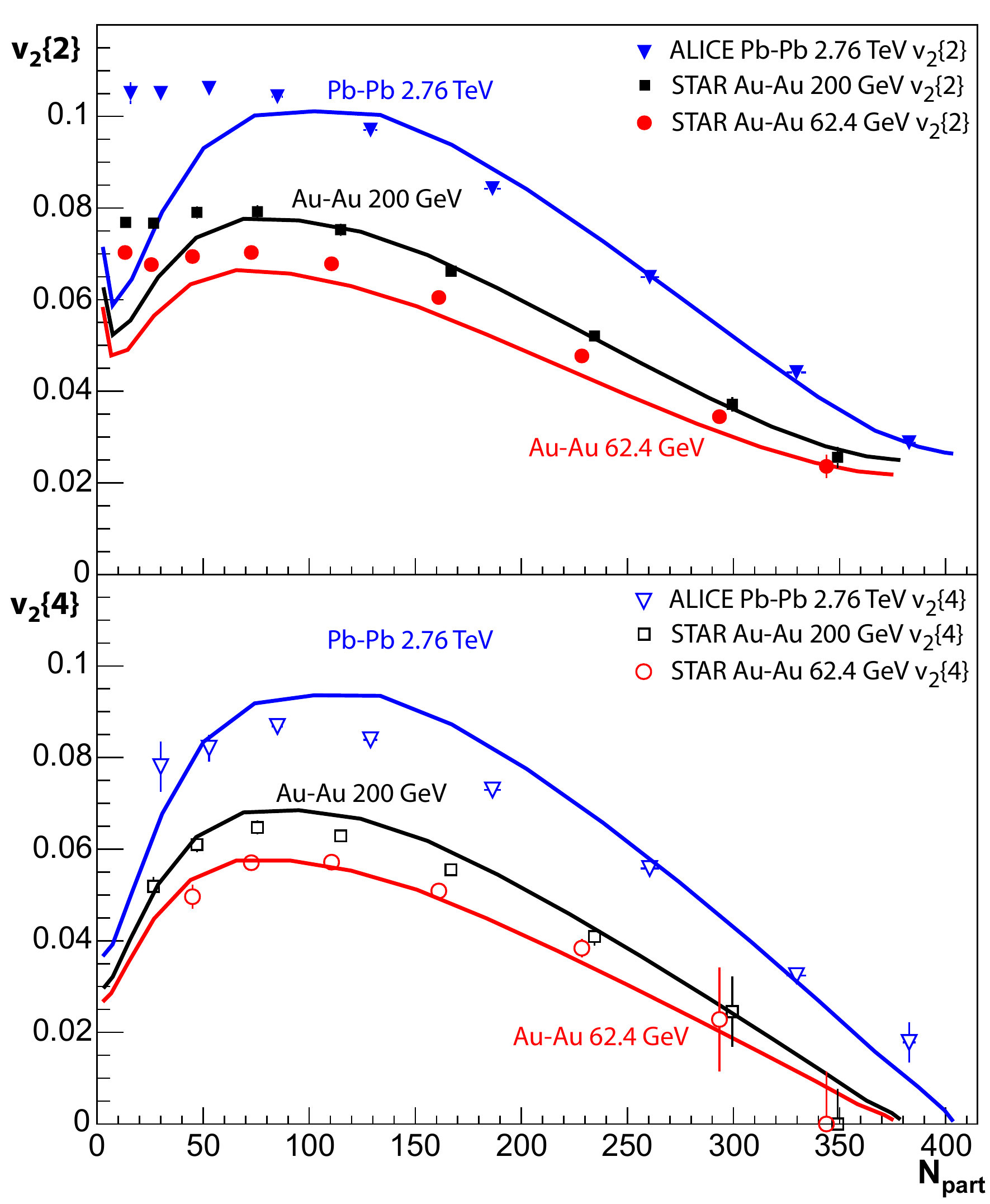}\hspace{2pc}%
\begin{minipage}[b]{14pc}\caption{\label{fig:v2}Measured $v_2\{2\}$ (top)  from STAR 
\cite{Pruthi:2011eq} and ALICE \cite{ALICE:2011ab,BilandzicThesis} compared to calculations using (\ref{eq:vnsigma}) and (\ref{eq:sigmaDef}).  Same for $v_2\{4\}$ (bottom) computed from (\ref{eq:vnRP}) in the blast wave.}
\end{minipage}
\end{figure}
\begin{figure}[h]
\begin{minipage}{14pc}
\includegraphics[width=14.3pc]{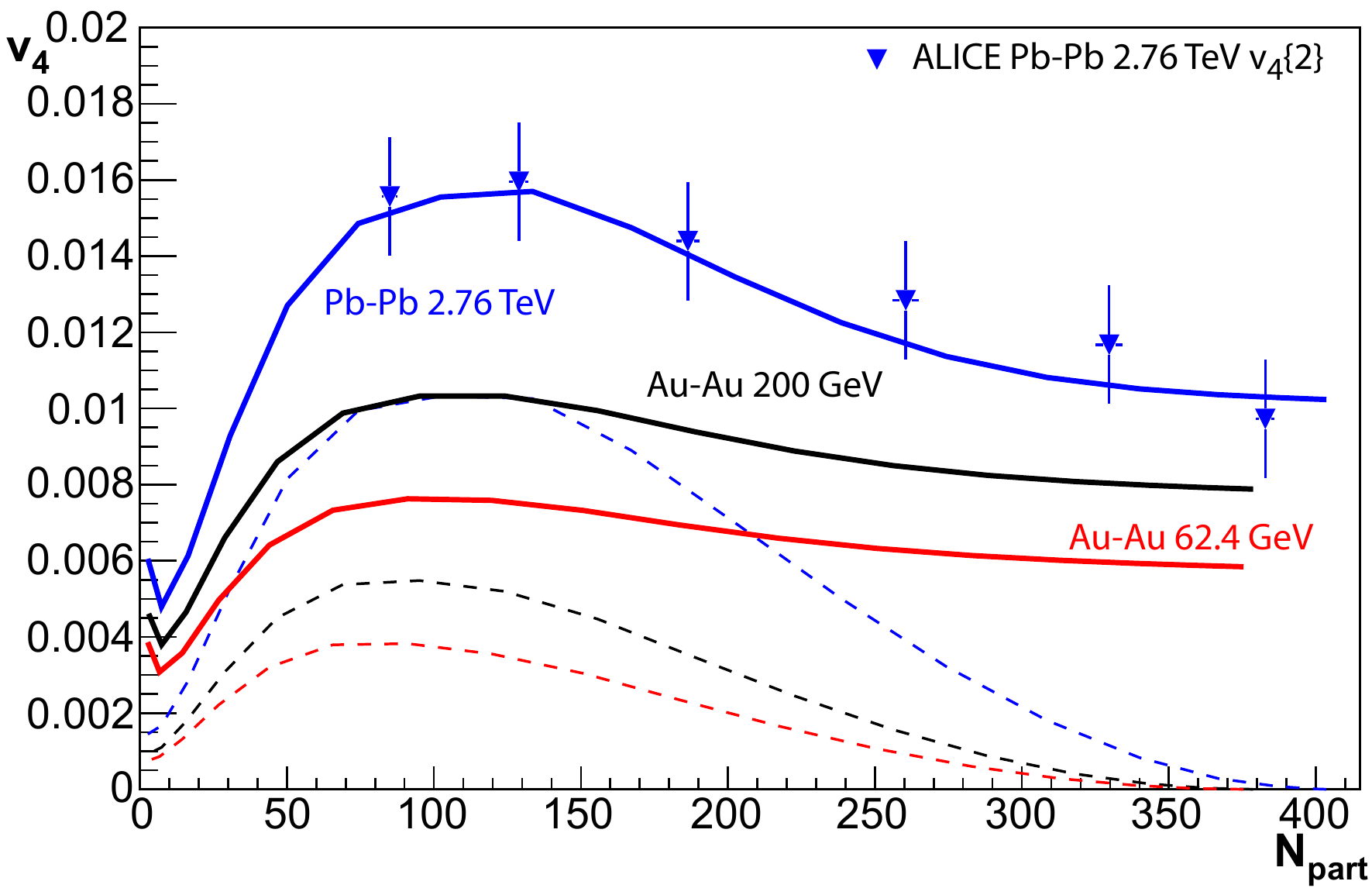}
\caption{\label{fig:v4}Same calculation as in Fig. \ref{fig:v2} but for the fourth order flow coefficients $v_4\{2\}$ (solid) and $v_4\{4\}$ (dashed). ALICE data  \cite{ALICE:2011ab}.}
\end{minipage}\hspace{2pc}%
\begin{minipage}{14pc}
\includegraphics[width=14pc]{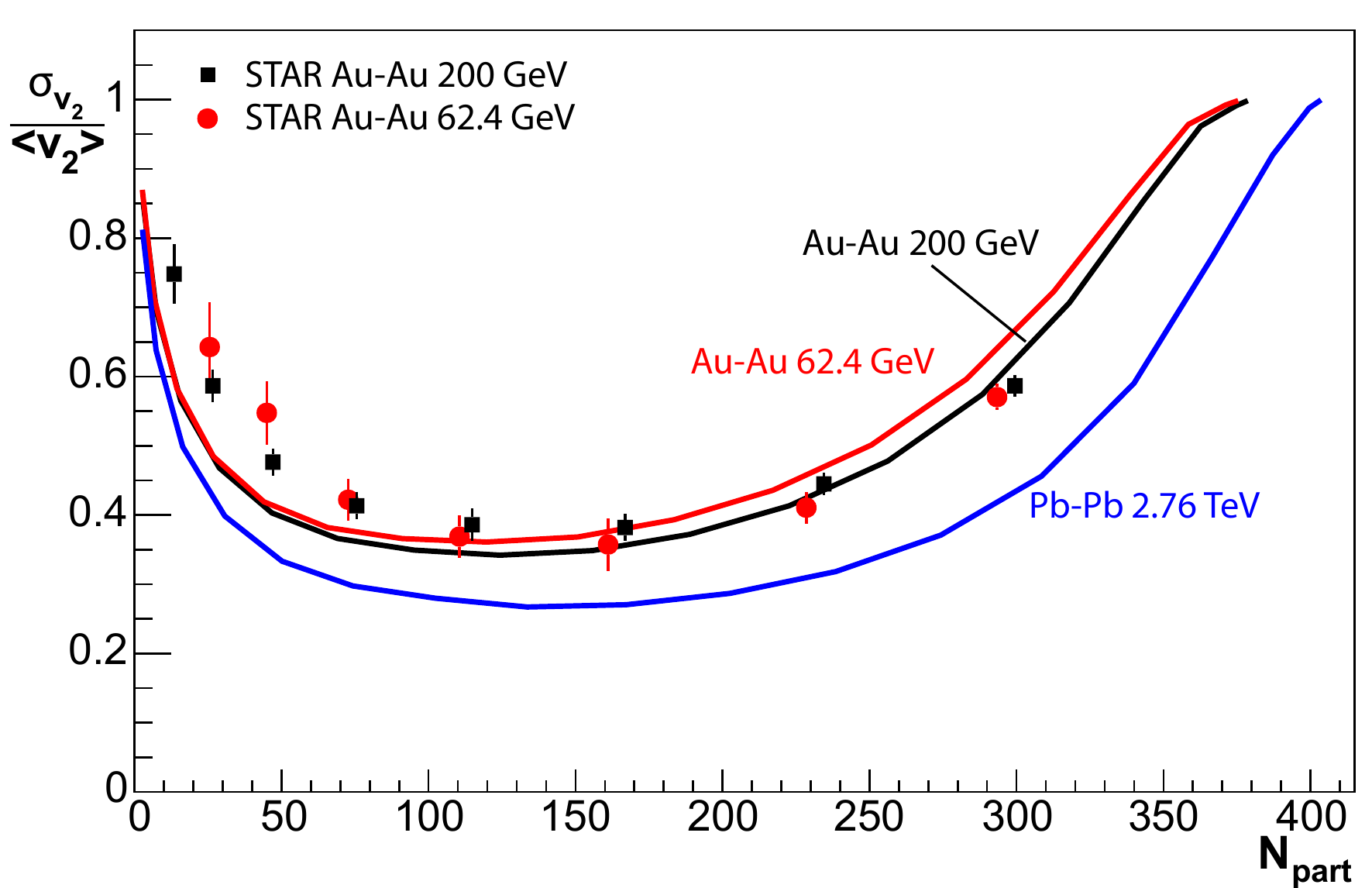}
\caption{\label{fig:v2Fano}Flow fluctuations in terms of the coefficient of variation for elliptic flow (\ref{eq:starFano}) compared to STAR data \cite{Pruthi:2011eq}.}
\end{minipage} 
\end{figure}

As with multiplicity fluctuations and $p_t$ fluctuations, the correlation strength $\R$ sets the scale for flow fluctuations. In our model, the increase in growth of flow fluctuations (\ref{eq:sigmaDef}) is coupled to the growth in the saturation scale $Q_s^2$ via (\ref{eq:CGCscale}). Comparisons of $v_2\{2\}$ and $v_2\{4\}$ to different collision energy measurements are shown in Fig. \ref{fig:v2}. 
It is also informative to make comparisons to $v_4$ as shown in Fig. \ref{fig:v4}. 
The effect of the change in blast wave parameters with collision energy is apparent from the separation of the solid lines in the bottom panel of Fig. \ref{fig:v2} and the dashed lines in Fig. \ref{fig:v4}. The importance of the Glasma contribution is similarly evident by the change is separation between $v_n\{4\}$ and $v_n\{2\}$ with
collision energy. 

STAR has also measured $v_2$ fluctuations in the form of
%
%
\begin{equation}\label{eq:starFano}
	\frac{\sigma_{v_n}}{\la v_n\ra} =
	\sqrt{\frac{v_n\{2\}^2 - v_n\{4\}^2}{v_n\{2\}^2 + v_n\{4\}^2}}
\end{equation}
resembling the so-called ``coefficient of variation'' defined as the standard deviation divided by the mean. Care should be taken here since the definition of $\sigma^2_{n}$, Eq.({\ref{eq:sigvn}), is not strictly the variance. In Fig. \ref{fig:v2Fano} we compare our calculation of (\ref{eq:starFano}) to measurement. Calculations at RHIC energies seem to agree reasonably well and we include the calculation for Pb+Pb 2.76 GeV as a prediction.
\begin{figure}[h]
\begin{minipage}{14pc}
\includegraphics[width=14.3pc]{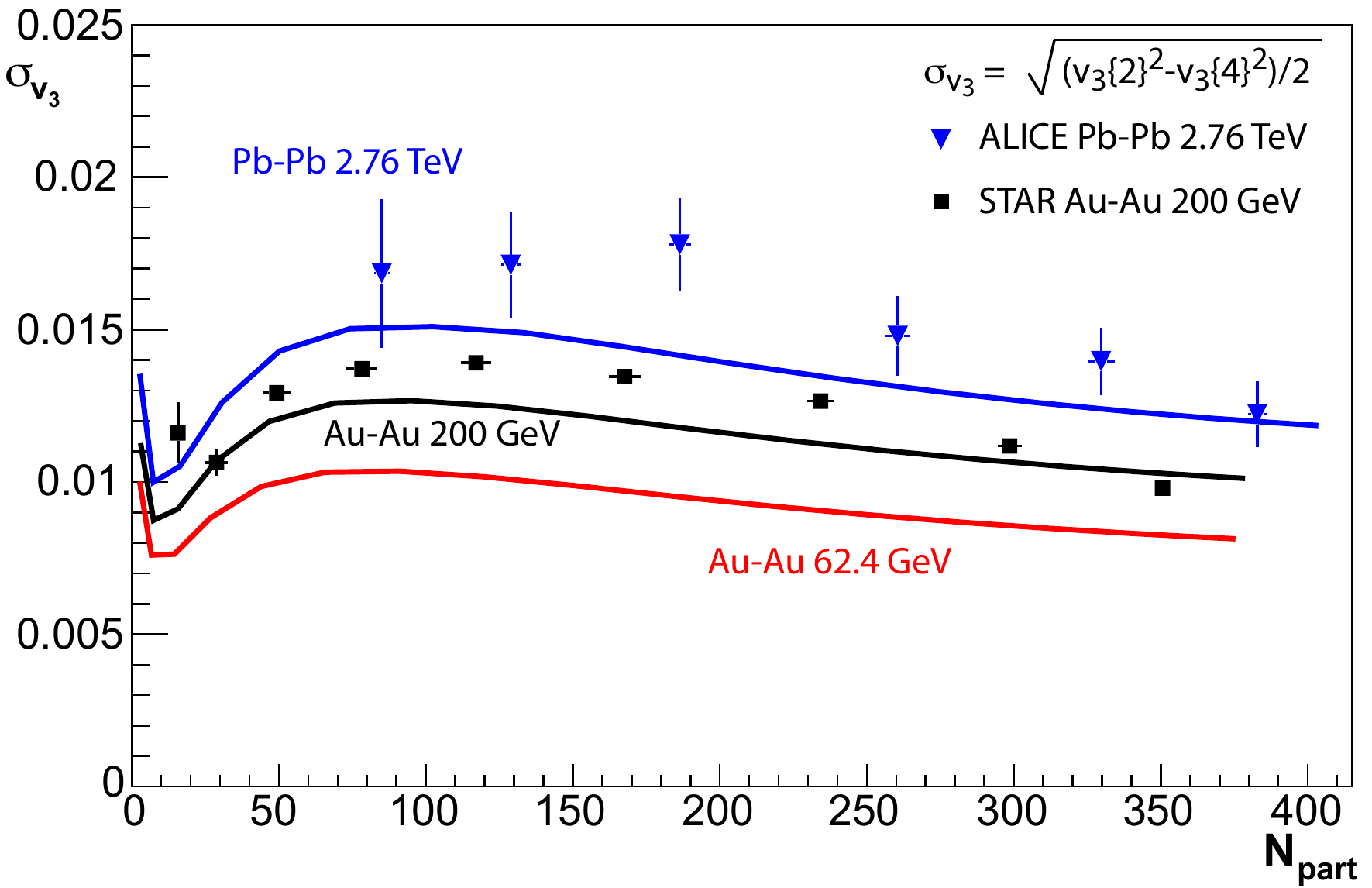}
\caption{\label{fig:sig_v3}Triangular flow fluctuations, $\sigma_{v_3}$, Eq.
(\ref{eq:sigmaDef}). Data points are computed from $v_3\{2\}$ and $v_3\{4\}$ measurements \cite{Sorensen:2011fb,ALICE:2011ab}.}
\end{minipage}\hspace{2pc}%
\begin{minipage}{14pc}
\includegraphics[width=14pc]{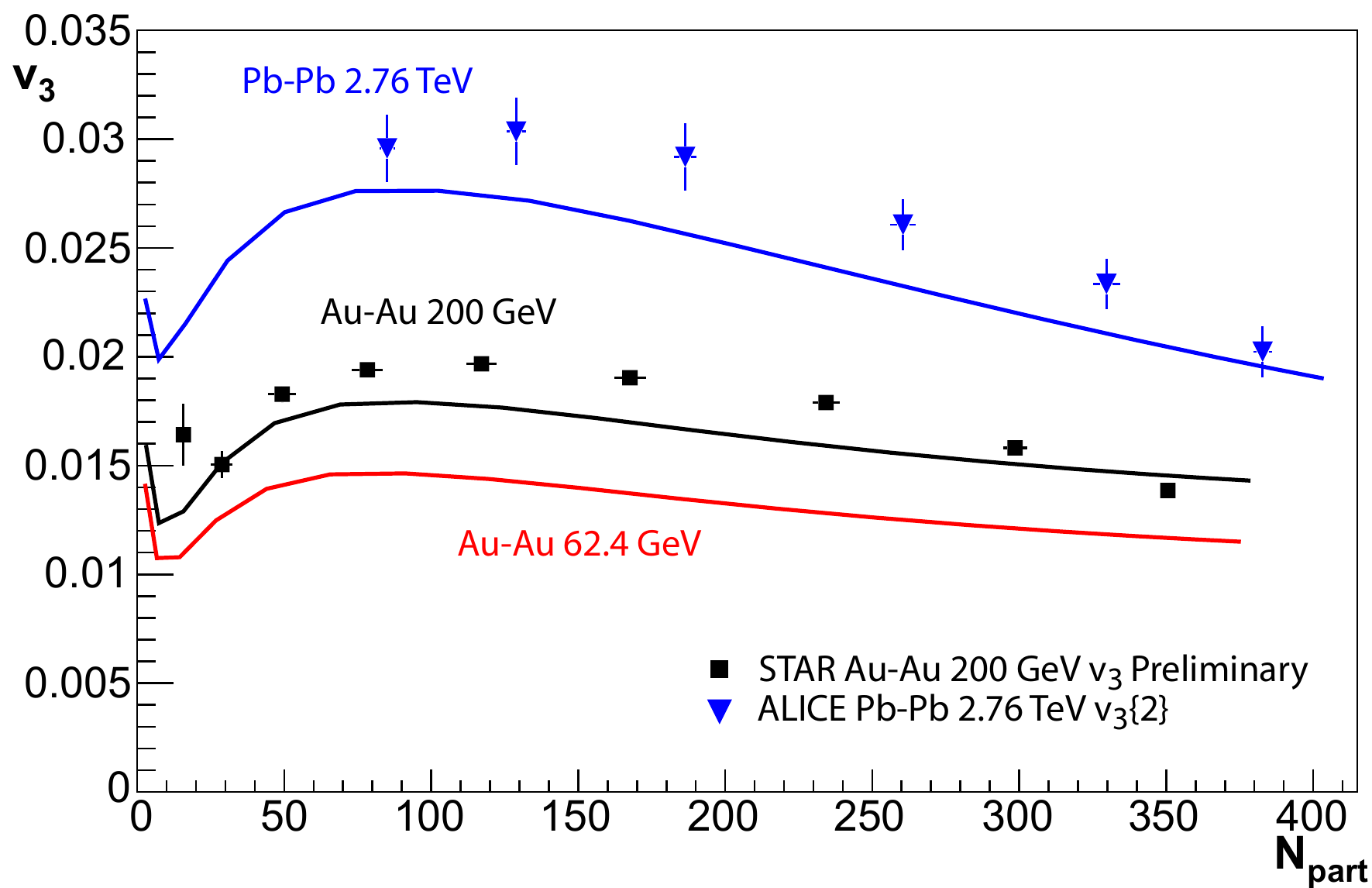}
\caption{\label{fig:v3}The triangular flow coefficient, $v_3\{2\}$ compared to
STAR and ALICE data \cite{Sorensen:2011fb,ALICE:2011ab}.}
\end{minipage} 
\end{figure}
\begin{figure}[h]
\includegraphics[width=14pc]{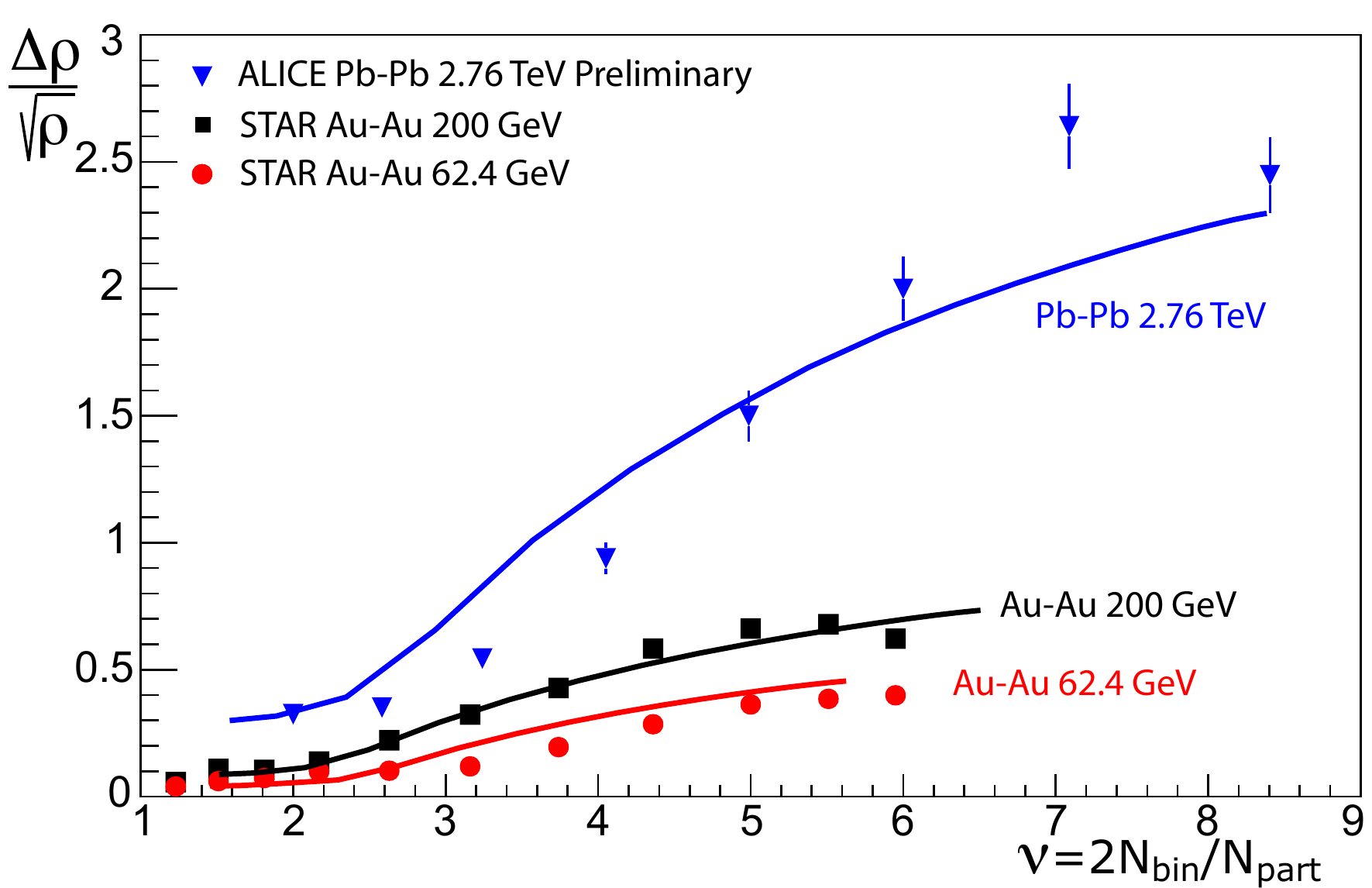}\hspace{2pc}%
\begin{minipage}[b]{14pc}\caption{\label{fig:AmpCompare}Near side ridge amplitude calculation from Glasma source correlations. Experimental data is from (STAR)
\cite{Daugherity:2006hz,Daugherity:2008zz,Daugherity:2008su} and (ALICE)
\cite{collaboration:2011um}.}
\end{minipage}
\end{figure}

In Fig. \ref{fig:sig_v3} we show $v_3$ fluctuations from Glasma. We further emphasize that  the energy dependence in Fig. \ref{fig:sig_v3} is in good accord with data, supporting the Glasma scaling with $Q_s^2$. The shape with centrality reflects contributions not only from Glasma, but also our parameterizations of the average reaction plane eccentricity $\varepsilon$ and our choice the flux tube distribution $\rho_{_{FT}}$.

To use (\ref{eq:sigmaDef}) and ({\ref{eq:sigvn}) to calculate $v_3\{2\}$, we must come to grips with the fact that our blast wave parametrization assumes $v_3\{4\}=0$. While this seems to be the case for the STAR measurements, ALICE has measured a non-zero $v_3\{4\}$ for Pb+Pb collisions at 2.76 TeV. To correct for this possible discrepancy, we can provide an ad-hoc parameterization of  $v_3\{4\}$ and use (\ref{eq:sigmaDef}) and
({\ref{eq:sigvn}) to calculate $v_3\{2\}$. Agreement shown in Fig. \ref{fig:v3} is
reasonable. As a preferable alternative, we compare our calculated $\sigma_{n}$ to fluctuations extracted from ALICE and STAR measurements in Fig. \ref{fig:sig_v3}.

The characterization of the ridge in terms of Fourier flow coefficients can prove a valuable check for models describing flow and its fluctuations with long range correlations. Relative contributions of flow and flow fluctuations to the ridge decoupled using Eq.({\ref{eq:FourierRidge}). Though it would be more informative to extract $\sigma_n$ directly from ridge measurements, we can compare the flow-subtracted near side ridge amplitude to integration of ({\ref{eq:MomCorr0}), suitably normalized to the mixed event background, see Refs.\cite{Gavin:2012if,Gavin:2008ev,Moschelli:2009tg} for details. 

Our calculation of the flow-subtracted near side ridge amplitude can be seen in Fig. \ref{fig:AmpCompare}. The dimensionless coefficient $\kappa$ in (\ref{eq:CGCscale}) is set by the change in amplitude of the 200 GeV Au+Au measurement and remains fixed for all other calculations. Like the other observables, the change in collision energy is determined by $\R$ and therefore $Q_s^2$. The agreement with data is quite reasonable. Commonality of the energy dependence between the ridge and flow fluctuations should not come as a surprise. Measurements of $v_2\{2\}$ at ALICE using particle pairs with and without a rapidity separation only show noticeable difference in peripheral collisions \cite{BilandzicThesis,Collaboration:2011yba}. The remaining difference between $v_2\{2\}$ and $v_2\{4\}$ suggests (\ref{eq:sigvn}) results from a long range correlation. However, long range correlation explanation of the energy dependence of multiplicity and momentum fluctuations does come as a surprise. Jets and resonance decays, for example, should have significant effects on these observables. Further investigation into the relative contributions of long and short range correlations can indicate if indeed long range correlations dominate fluctuation observables at these energies.

%
%
%
%
\section{\label{sec:discussion}Discussion}
In this work we study the connection between correlations and fluctuations of multiplicity, momentum, and harmonic flow coefficients. We first argue in Sec. \ref{sec:cff} that given the existence of genuine two-particle correlations in the hadron pair distribution, Eq. (\ref{eq:corrFunExp0}), multiplicity fluctuations, momentum fluctuations, flow fluctuations, and the ridge are all linked by a common scale $\R$. We then consider contributions to the genuine correlations due to fluctuating or lumpy initial parton distributions; correlated partons emerge from the same source/lump. 
We take the origin of initial state parton density lumps from Glasma flux tubes and present our model calculations in Sec. \ref{sec:results}. 

We focus on two important features of the data: the agreement with the change in collision energy and the link between the correlation and fluctuation observables. Long range correlations such as those measured in the ridge indicate that the correlations originate at early times in the collision. The connection between the ridge and flow fluctuations is evidence that the difference between $v_n\{4\}$ and $v_n\{2\}$ is also a long range phenomenon. 
Given that flow fluctuations result from the existence of parton density lumps in the initial state, one can compare the behavior of flow fluctuations to that of other consequences of the lumpy initial state such as multiplicity and momentum fluctuations.
The correlation strength $\R$ sets a common scale for all of these observables and controls much of the centrality and collision energy dependence. Theories of the initial state can therefore be constrained with measurements of $\R$. In this work we have chosen a CGC-Glasma based description of $\R$ and found reasonable agreement in the centrality and collision energy changes in all of the mentioned observables. This suggests that saturation based descriptions of the initial state are consistent with at least mid-central to central collisions.
\ack
This work was supported in part by the U.S. NSF grant PHY-0855369 (SG) and The Alliance Program of the Helmholtz Association (HA216/EMMI) (GM)

\section*{References}
\bibliographystyle{iopart-num}
\bibliography{ridge_fluctuations_refs}

\providecommand{\newblock}{}
\begin{thebibliography}{10}
\expandafter\ifx\csname url\endcsname\relax
  \def\url#1{{\tt #1}}\fi
\expandafter\ifx\csname urlprefix\endcsname\relax\def\urlprefix{URL }\fi
\providecommand{\eprint}[2][]{\url{#2}}

\bibitem{Gavin:2011gr}
Gavin S and Moschelli G 2012 {\em Phys.Rev.\/} {\bf C85} 014905
  (\textit{Preprint} \eprint{1107.3317})

\bibitem{Gavin:2012if}
Gavin S and Moschelli G 2012  (\textit{Preprint} \eprint{1205.1218})

\bibitem{Borghini:2000sa}
Borghini N, Dinh P~M and Ollitrault J~Y 2001 {\em Phys. Rev.\/} {\bf C63}
  054906 (\textit{Preprint} \eprint{nucl-th/0007063})

\bibitem{Borghini:2001vi}
Borghini N, Dinh P~M and Ollitrault J~Y 2001 {\em Phys.Rev.\/} {\bf C64} 054901
  (\textit{Preprint} \eprint{nucl-th/0105040})

\bibitem{Gavin:2008ev}
Gavin S, McLerran L and Moschelli G 2009 {\em Phys.Rev.\/} {\bf C79} 051902
  (\textit{Preprint} \eprint{0806.4718})

\bibitem{Moschelli:2009tg}
Moschelli G and Gavin S 2010 {\em Nucl.Phys.\/} {\bf A836} 43--58
  (\textit{Preprint} \eprint{0910.3590})

\bibitem{Voloshin:2003ud}
Voloshin S~A 2006 {\em Phys.Lett.\/} {\bf B632} 490--494 (\textit{Preprint}
  \eprint{nucl-th/0312065})

\bibitem{Pruneau:2007ua}
Pruneau C~A, Gavin S and Voloshin S~A 2008 {\em Nucl. Phys.\/} {\bf A802}
  107--121 (\textit{Preprint} \eprint{0711.1991})

\bibitem{Lindenbaum:2007ui}
Lindenbaum S~J and Longacre R~S 2007 {\em Eur. Phys. J.\/} {\bf C49} 767--782

\bibitem{Sorensen:2008bf}
Sorensen P 2008  (\textit{Preprint} \eprint{0811.2959})

\bibitem{Peitzmann:2009vj}
Peitzmann T 2009  (\textit{Preprint} \eprint{0903.5281})

\bibitem{Takahashi:2009na}
Takahashi J {\em et~al.\/} 2009 {\em Phys. Rev. Lett.\/} {\bf 103} 242301
  (\textit{Preprint} \eprint{0902.4870})

\bibitem{Andrade:2010xy}
Andrade R~P~G, Grassi F, Hama Y and Qian W~L 2010  (\textit{Preprint}
  \eprint{1008.4612})

\bibitem{Werner:2010aa}
Werner K, Karpenko I, Pierog T, Bleicher M and Mikhailov K 2010
  (\textit{Preprint} \eprint{1004.0805})

\bibitem{Pruneau:2002yf}
Pruneau C, Gavin S and Voloshin S 2002 {\em Phys. Rev.\/} {\bf C66} 044904
  (\textit{Preprint} \eprint{nucl-ex/0204011})

\bibitem{Adare:2008ns}
Adare A {\em et~al.\/} (PHENIX) 2008 {\em Phys. Rev.\/} {\bf C78} 044902
  (\textit{Preprint} \eprint{0805.1521})

\bibitem{Gelis:2009wh}
Gelis F, Lappi T and McLerran L 2009 {\em Nucl. Phys.\/} {\bf A828} 149--160
  (\textit{Preprint} \eprint{0905.3234})

\bibitem{Voloshin:1999yf}
Voloshin S, Koch V and Ritter H 1999 {\em Phys.Rev.\/} {\bf C60} 024901
  (\textit{Preprint} \eprint{nucl-th/9903060})

\bibitem{Voloshin:2001ei}
Voloshin S~A (STAR) 2001   591--596 (\textit{Preprint}
  \eprint{nucl-ex/0109006})

\bibitem{Adamova:2003pz}
Adamova D {\em et~al.\/} (CERES) 2003 {\em Nucl.Phys.\/} {\bf A727} 97--119
  (\textit{Preprint} \eprint{nucl-ex/0305002})

\bibitem{Adams:2005ka}
Adams J {\em et~al.\/} (STAR) 2005 {\em PHRVA,C72,044902.2005\/} {\bf C72}
  044902 (\textit{Preprint} \eprint{nucl-ex/0504031})

\bibitem{Gavin:2003cb}
Gavin S 2004 {\em Phys. Rev. Lett.\/} {\bf 92} 162301 (\textit{Preprint}
  \eprint{nucl-th/0308067})

\bibitem{Bilandzic:2010jr}
Bilandzic A, Snellings R and Voloshin S 2011 {\em Phys.Rev.\/} {\bf C83} 044913
  (\textit{Preprint} \eprint{1010.0233})

\bibitem{Voloshin:2008dg}
Voloshin S~A, Poskanzer A~M and Snellings R 2008  (\textit{Preprint}
  \eprint{0809.2949})

\bibitem{Sorensen:2009cz}
Sorensen P 2009  (\textit{Preprint} \eprint{0905.0174})

\bibitem{Voloshin:2007pc}
Voloshin S~A, Poskanzer A~M, Tang A and Wang G 2008 {\em Phys.Lett.\/} {\bf
  B659} 537--541 (\textit{Preprint} \eprint{0708.0800})

\bibitem{Dumitru:2008wn}
Dumitru A, Gelis F, McLerran L and Venugopalan R 2008 {\em Nucl. Phys.\/} {\bf
  A810} 91 (\textit{Preprint} \eprint{0804.3858})

\bibitem{Daugherity:2006hz}
Daugherity M (STAR) 2006 {\em PoS\/} {\bf CFRNC2006} 005 (\textit{Preprint}
  \eprint{nucl-ex/0611032})

\bibitem{Daugherity:2008zz}
Daugherity M~S 2008 {\em {Two-particle correlations in ultra relativistic heavy
  ion collisions}\/} Ph.D. thesis

\bibitem{Daugherity:2008su}
Daugherity M (STAR) 2008 {\em J. Phys.\/} {\bf G35} 104090 (\textit{Preprint}
  \eprint{0806.2121})

\bibitem{Kharzeev:2000ph}
Kharzeev D and Nardi M 2001 {\em Phys. Lett.\/} {\bf B507} 121--128
  (\textit{Preprint} \eprint{nucl-th/0012025})

\bibitem{Heckel_ALICEptFluc}
Heckel S (ALICE) 2011  (\textit{Preprint} \eprint{1107.4327})

\bibitem{Pruthi:2011eq}
Pruthi N~K (The STAR Collaboration) 2011  (\textit{Preprint}
  \eprint{1111.5637})

\bibitem{Gavin:2004dc}
Gavin S 2004 {\em J. Phys.\/} {\bf G30} S1385--S1388 (\textit{Preprint}
  \eprint{nucl-th/0404048})

\bibitem{ALICE:2011ab}
Aamodt K {\em et~al.\/} (ALICE Collaboration) 2011 {\em Phys.Rev.Lett.\/} {\bf
  107} 032301 (\textit{Preprint} \eprint{1105.3865})

\bibitem{BilandzicThesis}
Bilandzic A 2012 {\em {Anisotropic flow measurements in ALICE at the large
  hadron collider}\/} Ph.D. thesis {Utrecht University}

\bibitem{Sorensen:2011fb}
Sorensen P (STAR Collaboration) 2011 {\em J.Phys.G\/} {\bf G38} 124029 8 pages.
  Conference proceedings for Quark Matter 2011 (\textit{Preprint}
  \eprint{1110.0737})

\bibitem{collaboration:2011um}
Timmins A~R (ALICE) 2011  (\textit{Preprint} \eprint{1106.6057})

\bibitem{Collaboration:2011yba}
Snellings R 2011 {\em J. Phys.\/} {\bf G38} 124013 (\textit{Preprint}
  \eprint{1106.6284})

\end{thebibliography}

\end{document}